\documentclass[12pt]{article}
\usepackage{amsmath,amssymb}
\usepackage{float}
\usepackage{graphicx}
\usepackage[utf8x]{inputenc}
\usepackage{array}

\usepackage[T1]{fontenc}
\usepackage{mathptmx}

\usepackage{natbib}

\usepackage{fullpage}

\date{}

\begin{document}

\vspace*{5mm}


\begin{center}
\Large{
\textbf {Bee Cluster 3D: A system to monitor the temperature in a hive over time} }
\end{center}
\bigskip  
Olivier Besson\textsuperscript{1 *},
Herv\'e Mugnier\textsuperscript{2},
Aur\'elien Neveux\textsuperscript{2},
Ga\"elle Rey\textsuperscript{2},
Sully Vitry\textsuperscript{2}.
 \newline
\textbf{1} University of Neuch\^atel, CH-2000 Neuch\^atel, Switzerland
 \newline
\textbf{2} Mind Technology, F-74166 Saint-Julien-en-Genevois, France
\medskip  \newline
 All authors contributed equally to this work.
\medskip \newline
* Corresponding author: Olivier Besson  \newline
Institute of mathematics,
University of Neuch\^atel  \newline
11, Rue Emile Argand  \newline
CH-2000 Neuch\^atel,
Switzerland  \newline
E-mail address: olivier.besson@unine.ch

\section*{Abstract}
A new system, Bee Cluster 3D, allowing the study of the time evolution of the 3D temperature 
distribution in a bee hive is presented.
This system can be used to evaluate the cluster size and the location of the queen during winter. 
In summer, the device can be used to quantify the size of the brood nest and the breeding activity of the queen.
The system does not disturb the activity of the colony and can be used on any hive.
This electronic system was developed to be non-intrusive, miniaturized, and energy autonomous.

\section*{Keywords} 
3D temperature distribution time evolution in a hive, 
winter loss of colonies, 
brood nest time evolution,
bee cluster time evolution .

\section{Introduction}

Beekeepers know that the temperature distribution in a bee hive is an important parameter for colony development, 
see also \citep{abou,bar,basile,lensky,sar,tautz} and included references.

Various temperature measurement and observation methods have been developed, see, e.g.,\citep{dun}.
A detailed history of these observations is presented in \citep{meik} and the included references.
Recently, \citep{bar,niew,sta,becher} developed new data-logging systems.
These systems have few sensors, so they cannot provide a 3D representation of the hive temperature.

A new methodology to measure the real-time temperature distribution in a bee hive is presented in this paper. 
The sensor locations provide a 3D representation of this distribution and  enable investigation of:
\begin{itemize}
\item the density, the morphology and the size of the bee cluster in the hive;
\item the quality, the size, and the distribution of the brood nest;
\item the location of the queen in winter.
\end{itemize}
Moreover, the system can be used to evaluate the colony health status and:
\begin{itemize}
\item to know the temperature distribution in the hive as a function of time;
\item to have a primary indicator of winter loss of colonies, with a study of risk factors;
\item to study the influence of the materials used (e.g., wood, polyester) on heat loss.
\end{itemize}
Owing to the fine mesh on each frame, the Bee Cluster 3D system can measure thermoregulation in the colony,
providing a 3D thermal image with the following advantages.
\begin{itemize}
\item The system is non-intrusive in the sense that it is not necessary to open the hive to perform measurements.
\item The system is non-invasive because the sensors merge into the the wax frame.
\item The information is transmitted in real time.
\item The system is energy self-sufficient and is designed for field use. 
\end{itemize}
In general, the presented innovation covers the development of a multi-sensor modular system 
featuring real-time information about hive  development. This system is integrated into a "plug and play"
sensor architecture that feeds data to a network database.

This database can be connected to mathematics and vision software on a computer, smartphone, or
tablet PC or to a dedicated interface.

From a technical perspective, the main advantage of the Bee Cluster 3D system is the integration of the 
sensors into the wax frame, enabling natural development of the colony. This integrations allows the beekeeper
to work in the hive and avoid  disconnection and connection of the lines and manipulation of the frames. 
\section{Materials and Methods}
\subsection{System}

\begin{figure}[H]
\begin{center}
\includegraphics[width=12cm]{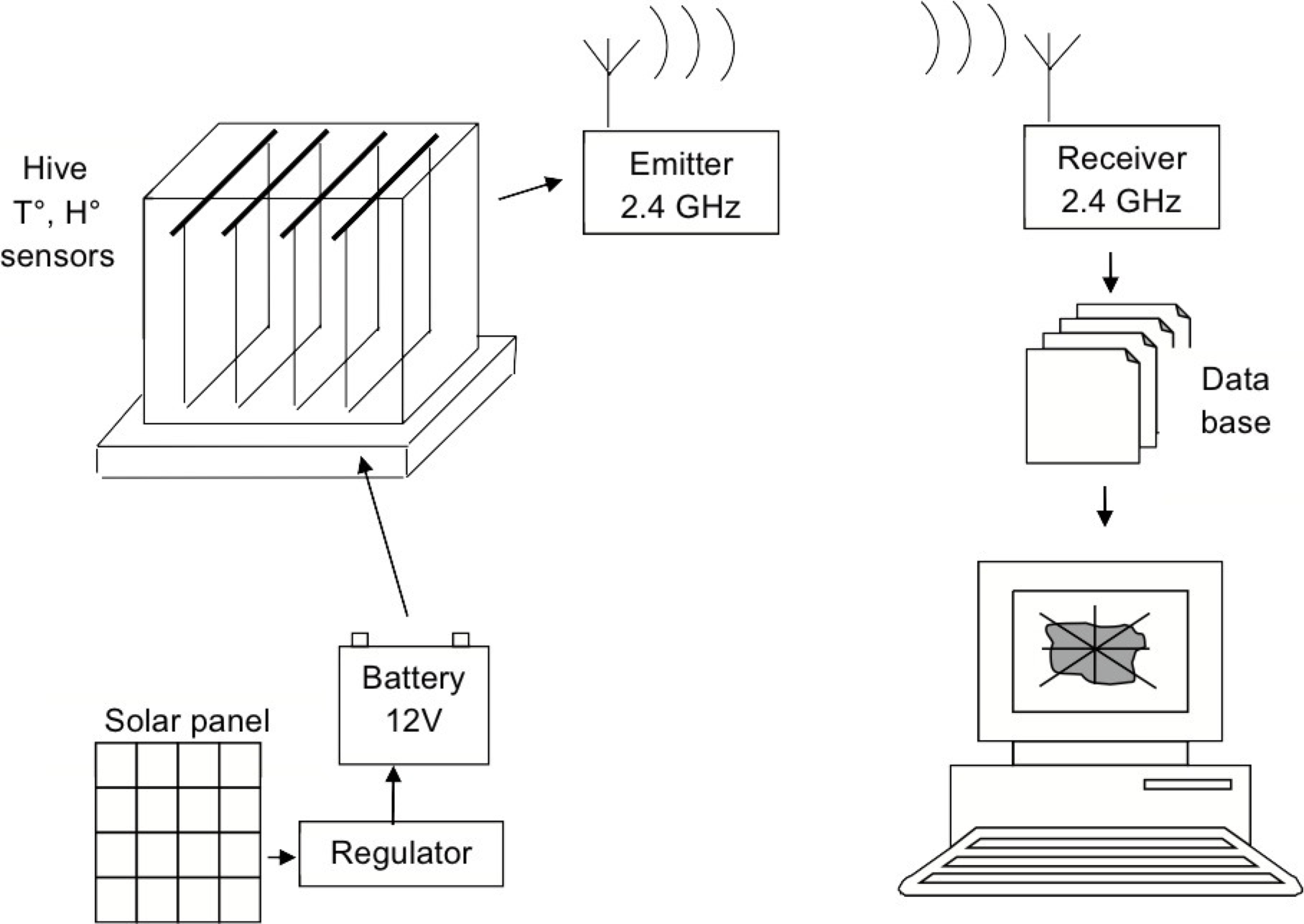}
\caption{{\bf Data collecting system.}}
\label{arch}
\end{center}
\end{figure}

An overview of the data acquisition system is  presented in figure \ref{arch}. The system consists of
the equipped hive,
the recording frames,
a power supply system,
a data transmitting/receiving component,
and a computer for the database.
\subsection{Equipment}

\begin{figure}[H]
\begin{center}
\includegraphics[width=12cm]{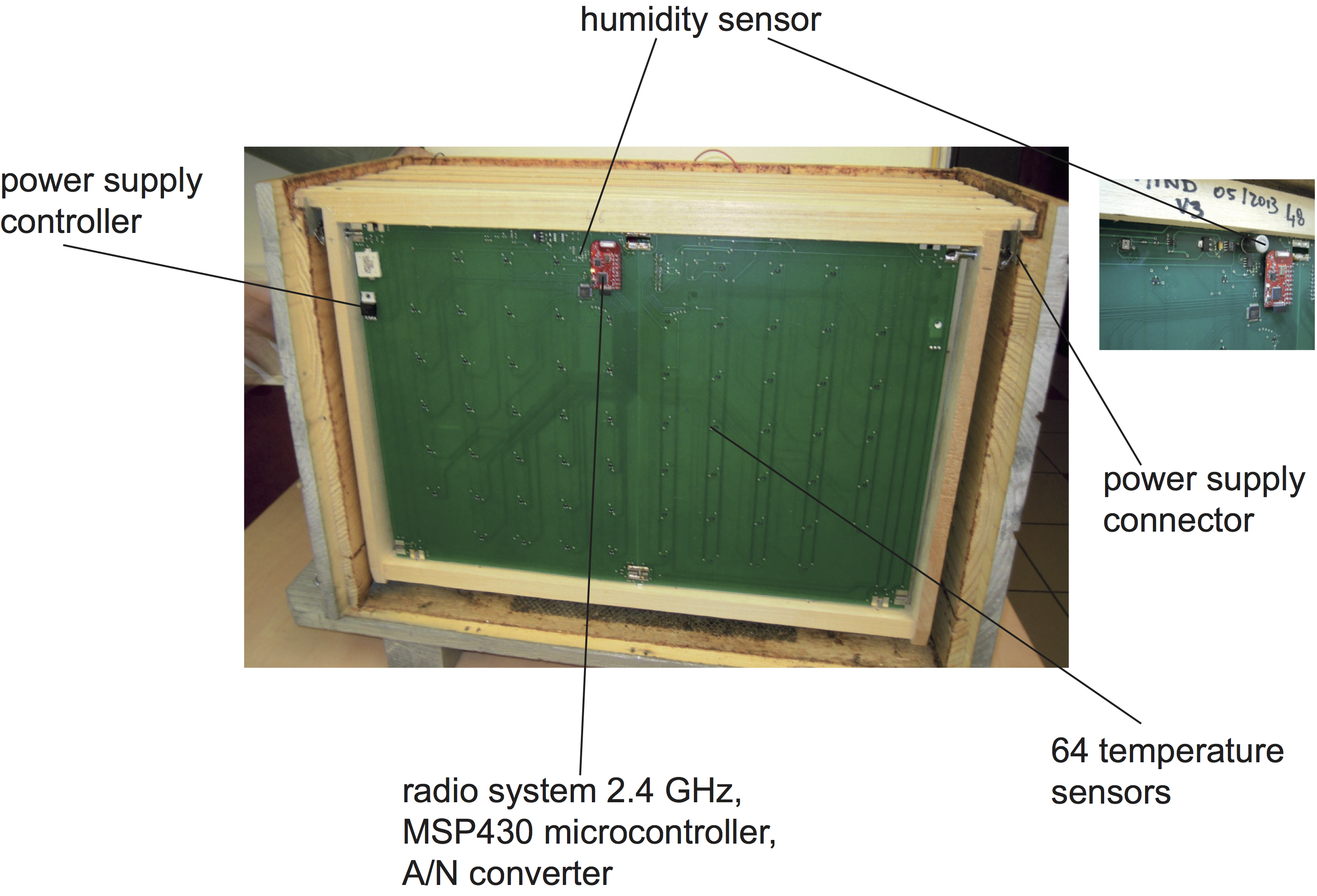}
\caption{{\bf Frame with electronic components.}}
\label{cadre_nu}
\end{center}
\end{figure}

The main data collecting components are the frames equipped with sensors. 
A frame without wax is shown in figure \ref{cadre_nu}.
It is equipped with
\begin{itemize}
\item 64 heat sensors with a spacing of 3 cm and an accuracy of $0.5^\circ C$.
\item 2 humidity sensors  with an accuracy of $0.5\%$.
\item A power management module.
\item A module for the electronic data processing and the radio system.
\item Electric power supplied by the connection to the frame rack.
\end{itemize}
From an energy perspective, the system was designed to be low power consuming.
The data recording is performed with a Raspberry Pi nano-computer.
\begin{itemize}
\item The energy consumption is 1 mW per frame.
\item The Raspberry Pi nano-computer consumes 5 W.
\item Data logging is performed every 5 minutes; this parameter may be modified.
\end{itemize}
To be fully operational, the instrumented frames must be built by bees, which
requires the use of wax comb foundations glued to each face of the electronic board
with acrylic glue.

\begin{figure}[H]
\begin{center}
\includegraphics[width=10cm]{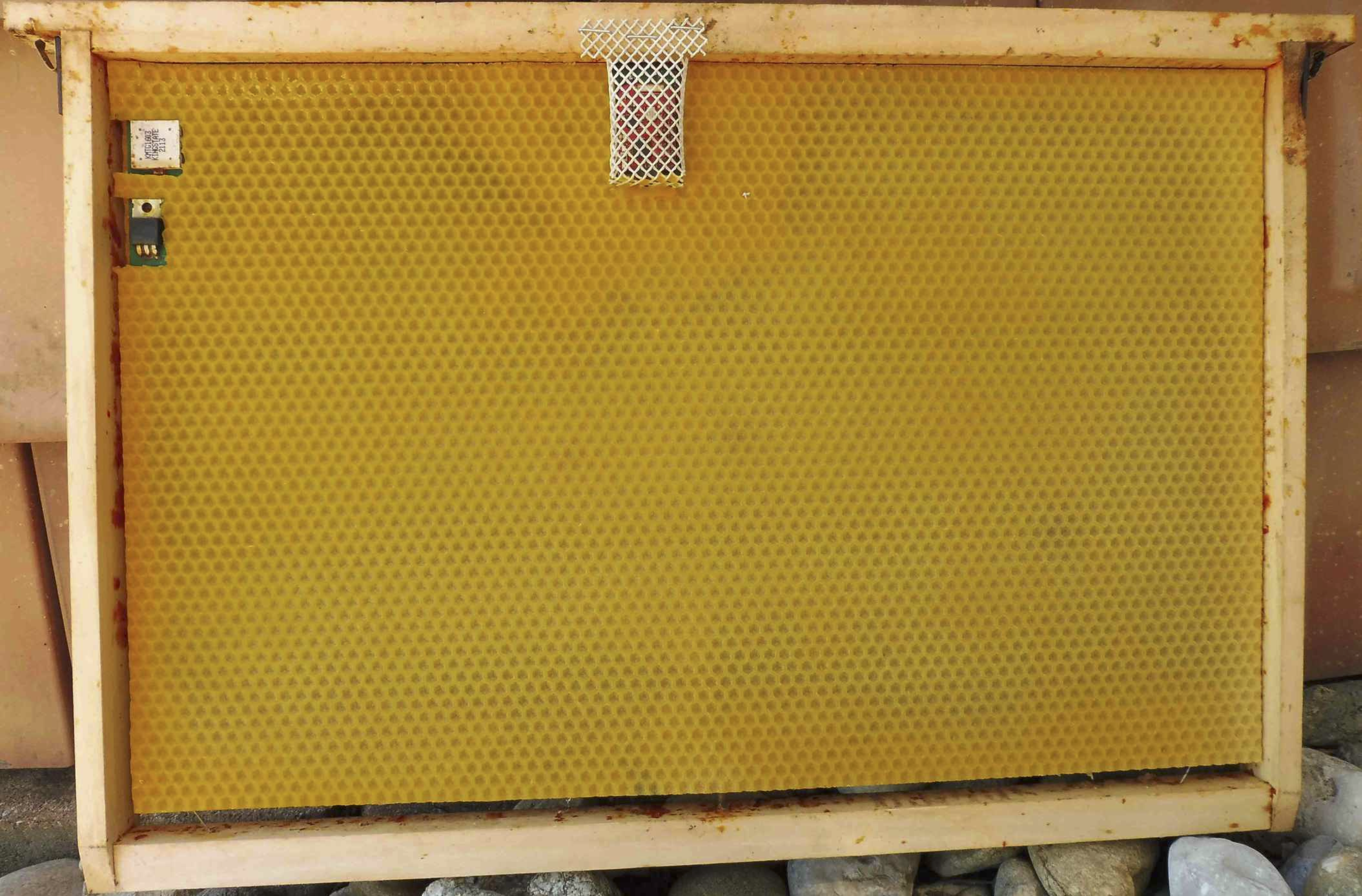}
\caption{{\bf Frame with a wax comb foundation.}}
\label{cire}
\end{center}
\end{figure}

\begin{figure}[H]
\begin{center}
\includegraphics[width=10cm]{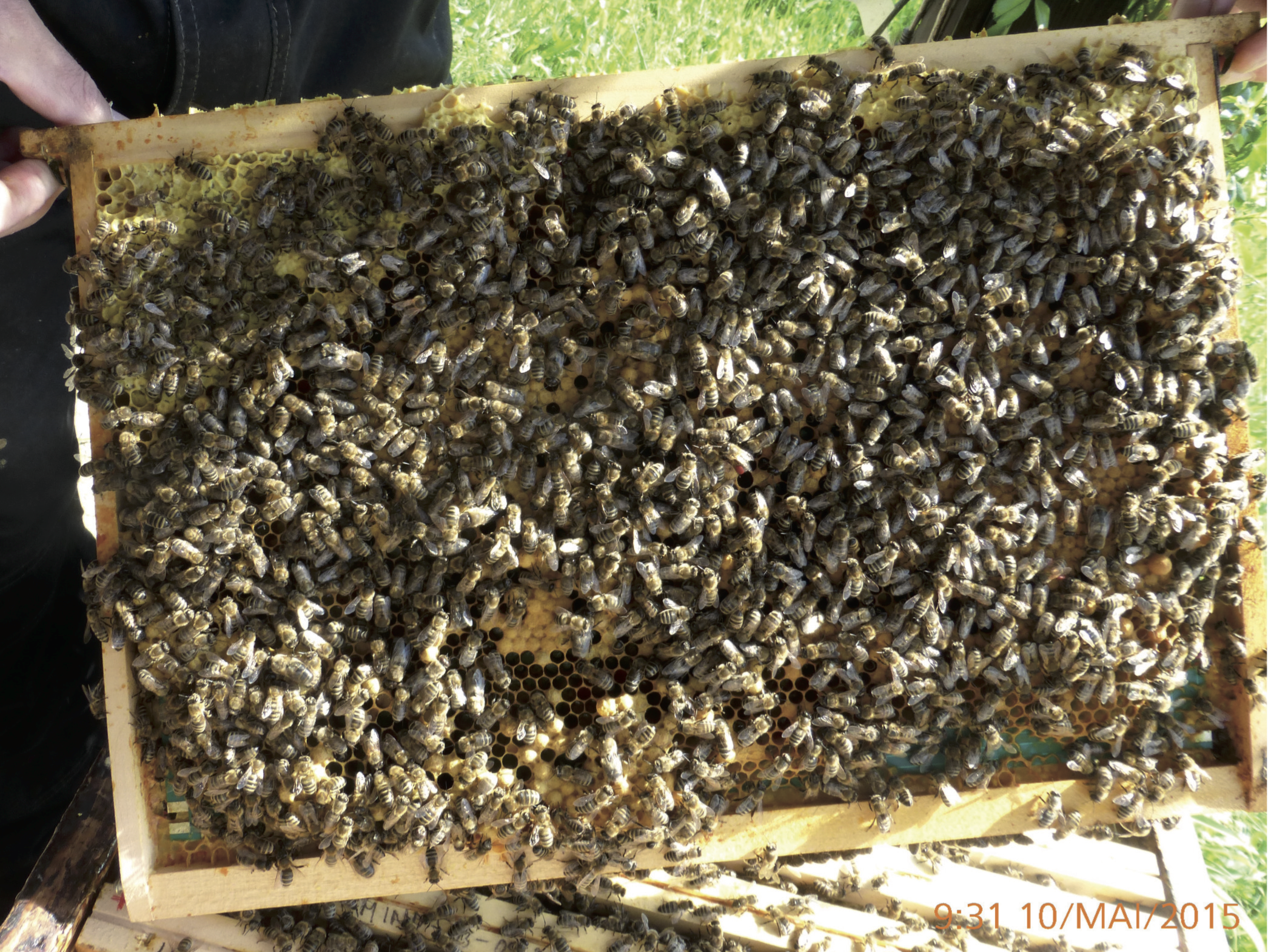}
\caption{{\bf Built frame.}}
\label{bati}
\end{center}
\end{figure}

A frame with a wax comb foundation and a built frame are presented in figure \ref{cire}, resp. \ref{bati}.
A non-metallic grid can be added to protect the radio system.
\section{Results}

The results presented in this section were obtained during the winter of 2014-2015 and 
the spring of 2015.
The observed hive was formed during autumn of 2014 as a 4-frame store. During the winter, the 
hive had 4 frames, and  new frames were introduced during the spring of 2015.

Our measurements enabled the following observations
\begin{itemize}
\item During the winter period, the maximum temperature remained greater than 30$^\circ$C
in a small, slowly moving region. This region corresponds to the location of the queen. 
\item During spring,  the activity of the queen resulted in an increase of the brood nest.
\item Our measurements provide an estimate of the evolution of the number brood cells.
\end{itemize}
Table \ref{tab:extrema} contains the temperature extrema observed from August 22, 2014 to
September 29, 2015, 
both in the hive and in a covered space near the hive.

This table shows that the maximum temperature remained above $33^\circ$C throughout  the winter.
As will be seen in section  \ref{hiver}, a small volume in the hive remained warmer than 
$30^\circ$C.
\begin{table}[H]
  \centering
  \small{
    \begin{tabular}{|c|c|c|c||c|c|}
    \hline
    \multicolumn{2}{|c|}{Date} & \multicolumn{2}{|c||}{Hive temperature $\mbox{ } ^\circ C$} &
     \multicolumn{2}{|c|}{External temperature $\mbox{ }^\circ C$} \\ \hline
    From & To & Minimum & Maximum & Minimum & Maximum \\ \hline
    22/08/2014 & 31/08/2014 & 15.90 & 37.60 &       &    \\ \hline
    31/08/2014 & 05/09/2014 & 17.80 & 40.20 &       &    \\ \hline
    21/09/2014 & 25/09/2014 & 12.60 & 36.50 &       &    \\ \hline
    30/09/2014 & 08/10/2014 & 17.70 & 36.80 & 13.30 & 26.30 \\ \hline
    08/10/2014 & 16/10/2014 & 14.30 & 35.90 & 12.00 & 25.30 \\ \hline
    08/12/2014 & 14/12/2014 & 0.10  & 33.40 & 1.10  & 25.00 \\ \hline
    14/12/2014 & 21/12/2014 & 1.30  & 35.00 & 2.70  & 15.50 \\ \hline
    29/12/2014 & 04/01/2015 & -4.00 & 33.30 & -2.60 & 11.80 \\ \hline
    11/01/2015 & 18/01/2015 & -2.60 & 33.50 & -0.90 & 13.90 \\ \hline
    18/01/2015 & 25/01/2015 & -0.60 & 33.20 & -4.70 & 16.10 \\ \hline
    26/01/2015 & 31/01/2015 & -1.60 & 33.70 & 0.40  & 6.30  \\ \hline
    31/01/2015 & 06/02/2015 & -3.10 & 33.00 & -1.60 & 8.10 \\ \hline
    08/02/2015 & 14/02/2015 & -3.70 & 33.60 & -1.20 & 11.10 \\ \hline
    14/02/2015 & 20/02/2015 & -1.90 & 34.50 & 0.10  & 9.70 \\ \hline
    02/03/2015 & 06/03/2015 & 1.30  & 35.30 &       &    \\ \hline
    06/03/2015 & 15/03/2015 & -1.20 & 35.80 &       &    \\ \hline
    15/03/2015 & 23/03/2015 & 3.50  & 36.70 &       &    \\ \hline
    23/03/2015 & 30/03/2015 & 5.80  & 37.70 & 1.90  & 18.30 \\ \hline
    30/03/2015 & 06/04/2015 & 6.20  & 37.70 & 2.10  & 18.70 \\ \hline
    06/04/2015 & 13/04/2015 & 6.20  & 42.30 & 2.70  & 23.80 \\ \hline
    16/04/2015 & 24/04/2015 & 14.20 & 43.00 & 6.00  & 24.20 \\ \hline
    24/04/2015 & 02/05/2015 & 22.00 & 42.40 &       &      \\ \hline
    02/05/2015 & 10/05/2015 & 27.00 & 37.50 &       &      \\ \hline
    17/05/2015 & 22/05/2015 & 22.50 & 38.00 &       &      \\ \hline
    22/05/2015 & 10/06/2015 & 23.20 & 37.90 &       &      \\ \hline
    10/06/2015 & 19/06/2015 & 25.20 & 37.20 &       &      \\ \hline
    19/06/2015 & 21/06/2015 & 26.50 & 37.30 &       &      \\ \hline
    29/06/2015 & 02/07/2015 & 21.10 & 37.80 &       &      \\ \hline
    23/07/2015 & 24/07/2015 & 15.40 & 37.90 &       &      \\ \hline
    28/07/2015 & 03/08/2015 & 23.50 & 36.70 &       &      \\ \hline
    03/08/2015 & 05/08/2015 & 18.60 & 37.90 &       &      \\ \hline
    26/08/2015 & 03/09/2015 & 11.80 & 37.80 &       &      \\ \hline
    03/09/2015 & 09/09/2015 & 14.20 & 39.40 &       &      \\ \hline
    26/09/2015 & 29/09/2015 & 15.80 & 38.30 &       &      \\ \hline
    \end{tabular}%
    }
  \caption{\textbf{Temperature extrema}}
  \label{tab:extrema}
\end{table}

Detailed results are presented in the following sections. In part  \ref{hiver}, the results obtained during the
winter of 2014-2015 are given. Section \ref{couvain} is devoted to the spring of 2015. 
The study of the brood evolution is analyzed via graphic localization and
by estimating the number of brood cells on each frame in the hive.

\subsection{Winter temperature distribution in a hive}
\label{hiver}

As already mentioned, throughout the winter, the maximum temperature in the hive remains above 30$^\circ$C
in the small region where the queen is located. However, the temperature in other parts of the hive can be below 0$^\circ$C,
as shown in table \ref{tab:extrema}.

The three following figures show the temperature distributions in the hive on January 18, 2015 at 6 AM.
Note that the temperature extrema in the hive are -2.6$^\circ$C and  33.1$^\circ$C.
All the winter observations are similar to those presented below.
As usual during winter, the bee cluster is located in the front of the hive.

\begin{figure}[H]
\begin{center}
\includegraphics[width=12cm]{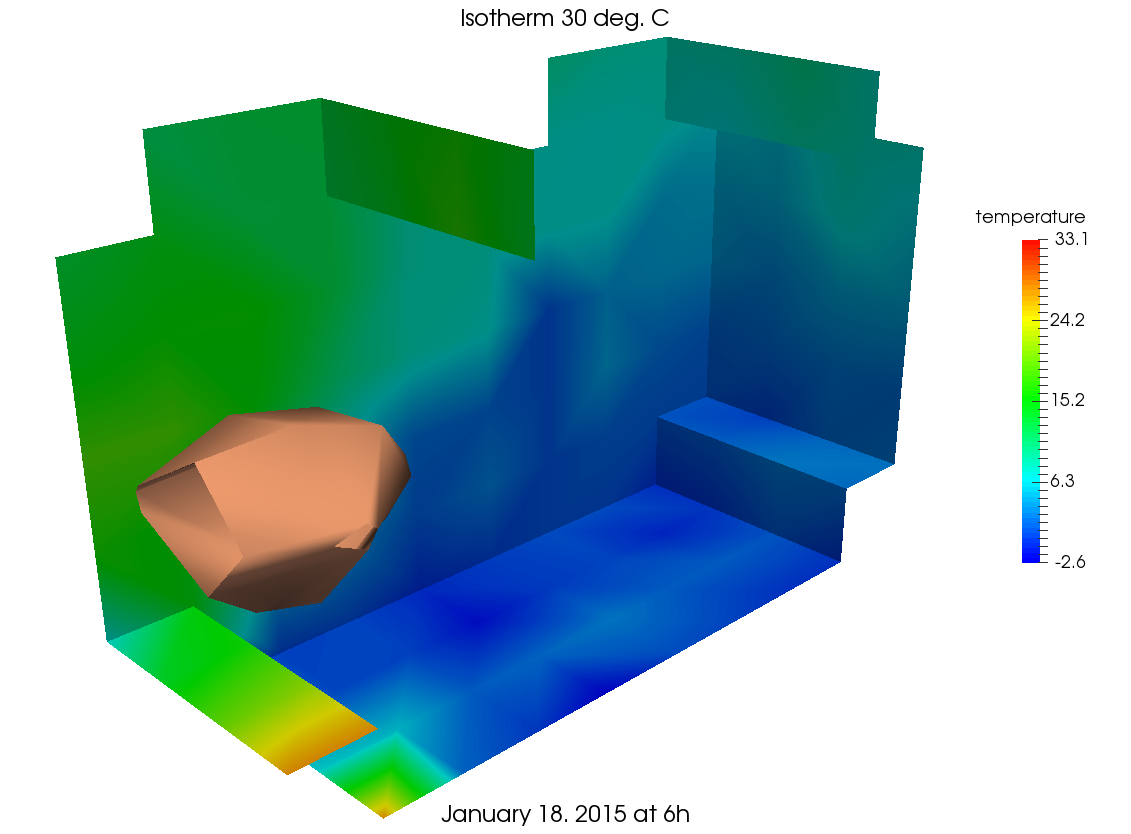}
\caption{{\bf 30$^\circ$C isotherm surface in the hive.}}
\label{iso30}
\end{center}
\end{figure}

\begin{figure}[H]
\begin{center}
\includegraphics[width=12cm]{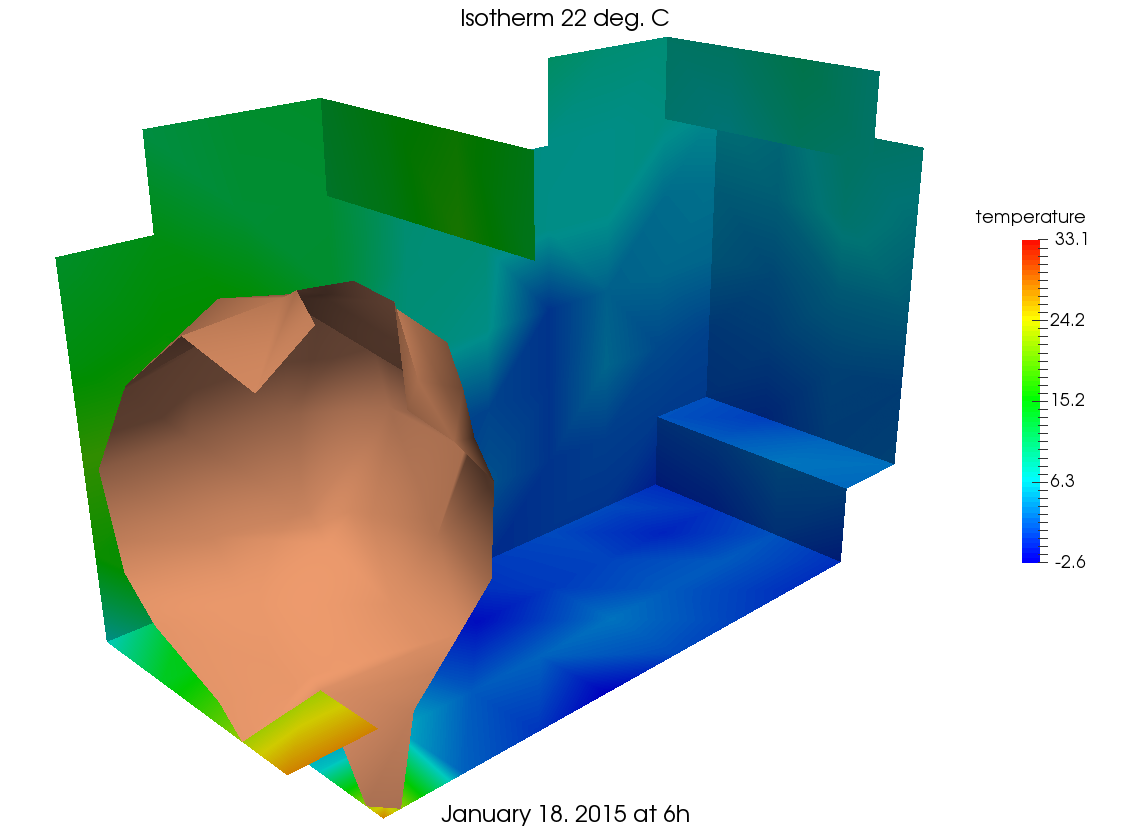}
\caption{{\bf 22$^\circ$C isotherm surface in the hive.}}
\label{iso22}
\end{center}
\end{figure}

\begin{figure}[H]
\begin{center}
\includegraphics[width=12cm]{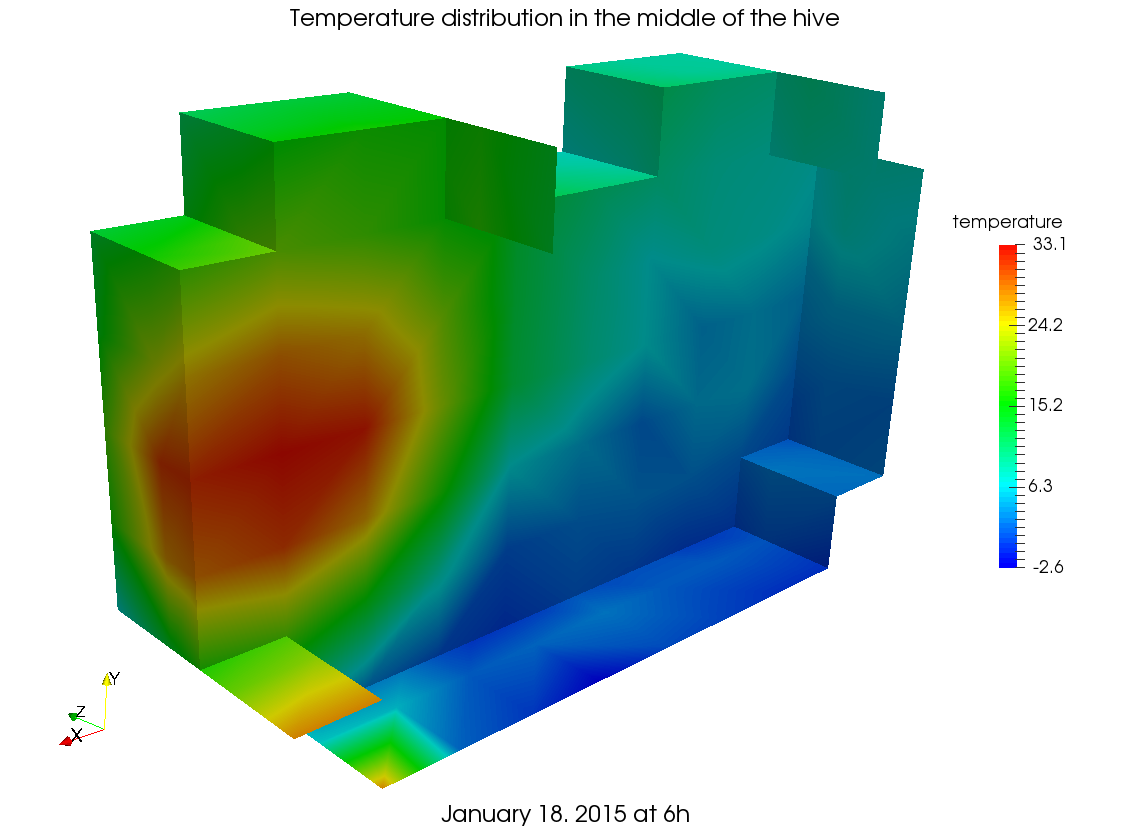}
\caption{{\bf Temperature distribution in the center of the hive.}}
\label{coupe18}
\end{center}
\end{figure}

Figure \ref{iso30} shows the location of the 30$^\circ$C isotherm surface in the hive; 
the queen is located inside this surface.
Figure \ref{iso22} shows the 22$^\circ$C isotherm surface. 
The bee cluster is included in this surface.
Finally, Figure \ref{coupe18} shows the temperature  distribution  of a slice in the middle of the hive.
\subsection{Brood development}
\label{couvain}

Brood development in hives occurs in the spring. The Bee Cluster 3D system can be used to precisely follow  this evolution
via a graphical representation of the volume of the brood nest and by quantifying the number
of brood cells on each frame.

In this section, observations made on the hive equipped with the Bee Cluster 3D system are presented for April 17, 2015.
During this period, the hive had seven frames, numbered 1 to 7.

\begin{figure}[H]
\begin{center}
\includegraphics[width=12cm]{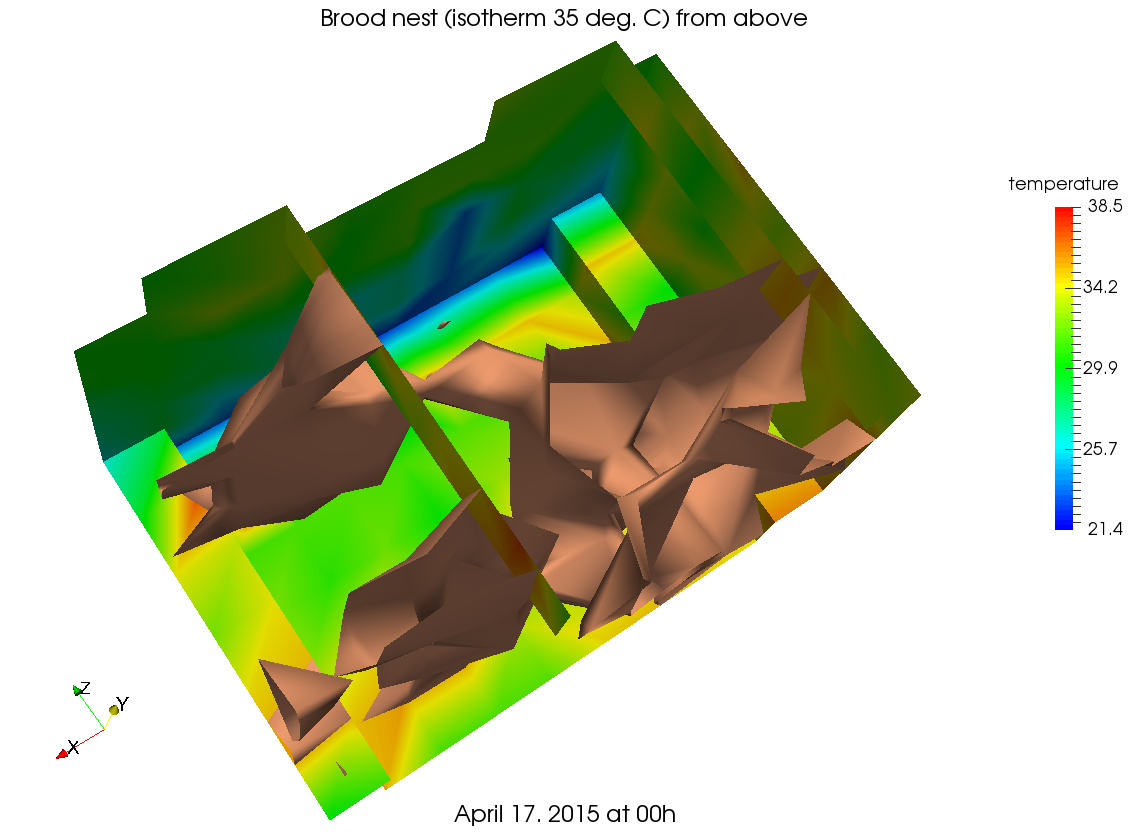}
\caption{{\bf Brood nest from above.}}
\label{nid_dessus}
\end{center}
\end{figure}

\begin{figure}[H]
\begin{center}
\includegraphics[width=12cm]{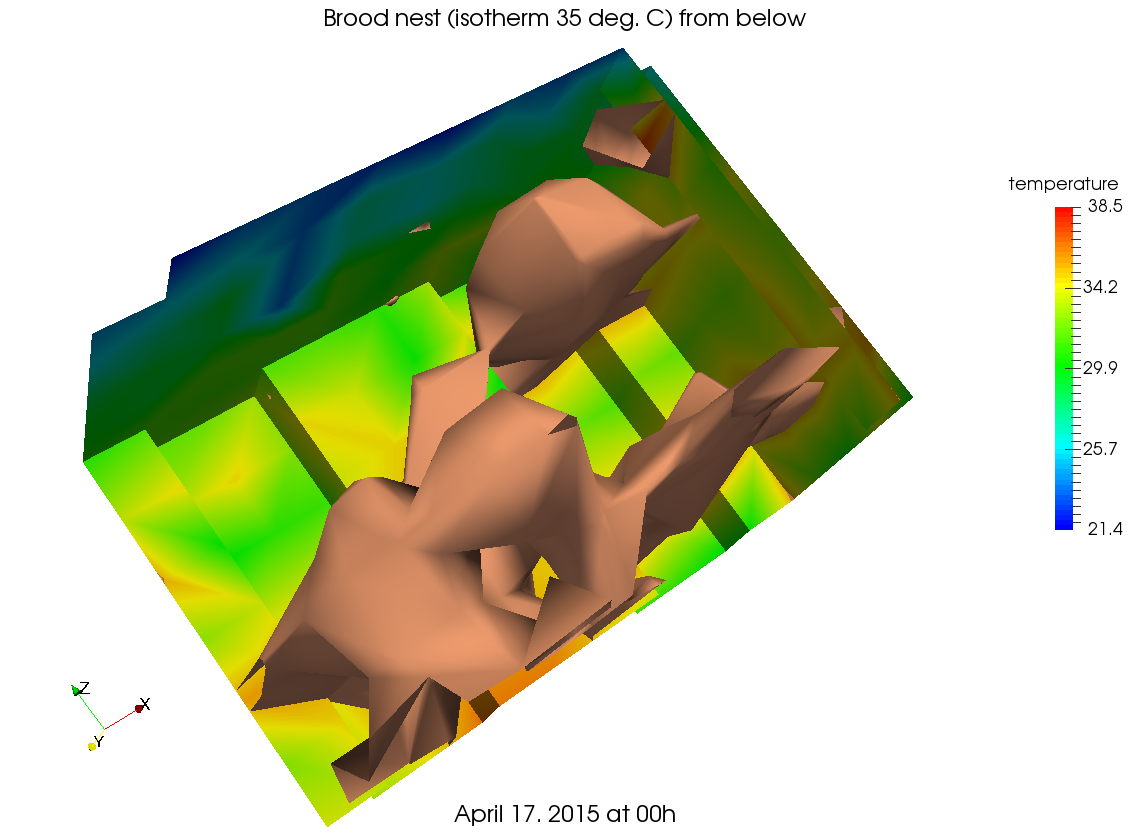}
\caption{{\bf Brood nest from below.}}
\label{nid_dessous}
\end{center}
\end{figure}

Figures \ref{nid_dessus} and \ref{nid_dessous}  show images of the brood nest from above and below, respectively.
These representations are the 35$^\circ$C isotherm surface in the hive.
As already mentioned, the total number of brood cells in the hive can be estimated in real time.
For this process, the brood quantity on each frame is obtained along with an estimate of the area of the
surface delimited  by an $T^\circ$C-isotherm curve, where $T$ is given by the user. 
Then, the number of brood cells is given by the formula
\begin{equation}
\label{nbcel}
N = 2 \frac{A}{A_c}
\end{equation}
where $N$ is the number of cells, 
$A$ is the area occupied by the brood cells, 
and $A_c$ is the area of one cell. $A_c$ is given by
\begin{equation}
A_c = \frac{\sqrt{3}}{2}d_i^2 = \frac{3\sqrt{3}}{8}d_e^2
\end{equation}
where $d_i$ is the diameter of the inscribed circle, and $d_e$ is the circumscribed circle of a cell.
The factor $2$ in formula \eqref{nbcel} accounts for the fact that a frame has two faces.

\begin{figure}[H]
\begin{center}
\includegraphics[width=12cm]{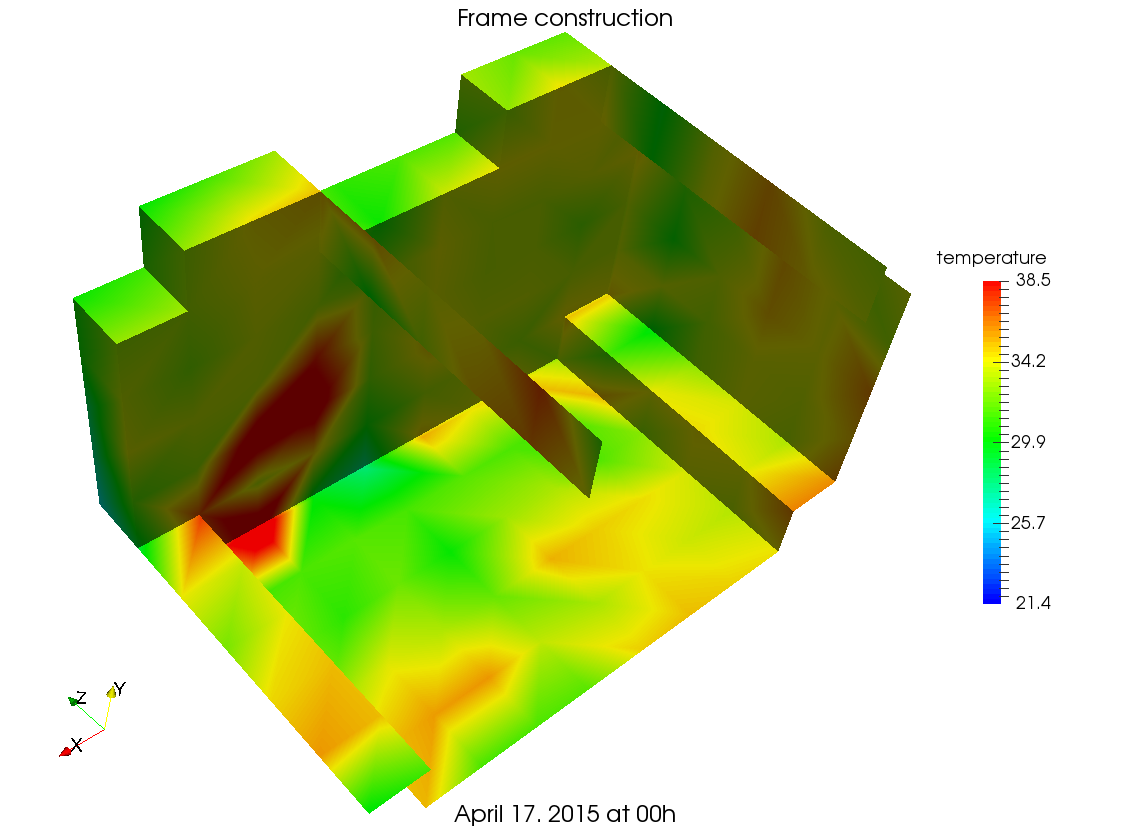}
\caption{{\bf Frame 6 under construction.}}
\label{construc}
\end{center}
\end{figure}

In the following examples, we use the following parameters:
\begin{itemize}
\item $T = 35.5^\circ C$
\item $d_i=5.3 mm$
\end{itemize}
The number of brood cells on each frame, calculated with the preceding method, is given in table  \ref{tab:nbcel}.
\begin{table}[H]
  \centering
    \begin{tabular}{|c|c|}
    \hline
    Frame & Number of brood cells \\ \hline
    1 & 890  \\ \hline 
    2 & 1018 \\ \hline 
    3 & 1578 \\ \hline 
    4 & 29  \\ \hline 
    5 & 550 \\ \hline 
    6 & 0 \\ \hline 
    7 & 0  \\ \hline 
    \end{tabular}%
  \caption{\textbf{Number of cells on the frames}}
  \label{tab:nbcel}
\end{table}
Note that frame 6 has no brood cells because this frame was introduced one day previously. Figure \ref{construc} shows the activity of the bees building this frame.
Note that the maximum temperature on this frame is 42$^\circ$C.
The following figures show the temperature distributions on the frames and the 35.5$^\circ$C isotherm curve.
\begin{figure}[H]
\begin{center}
\includegraphics[width=12cm]{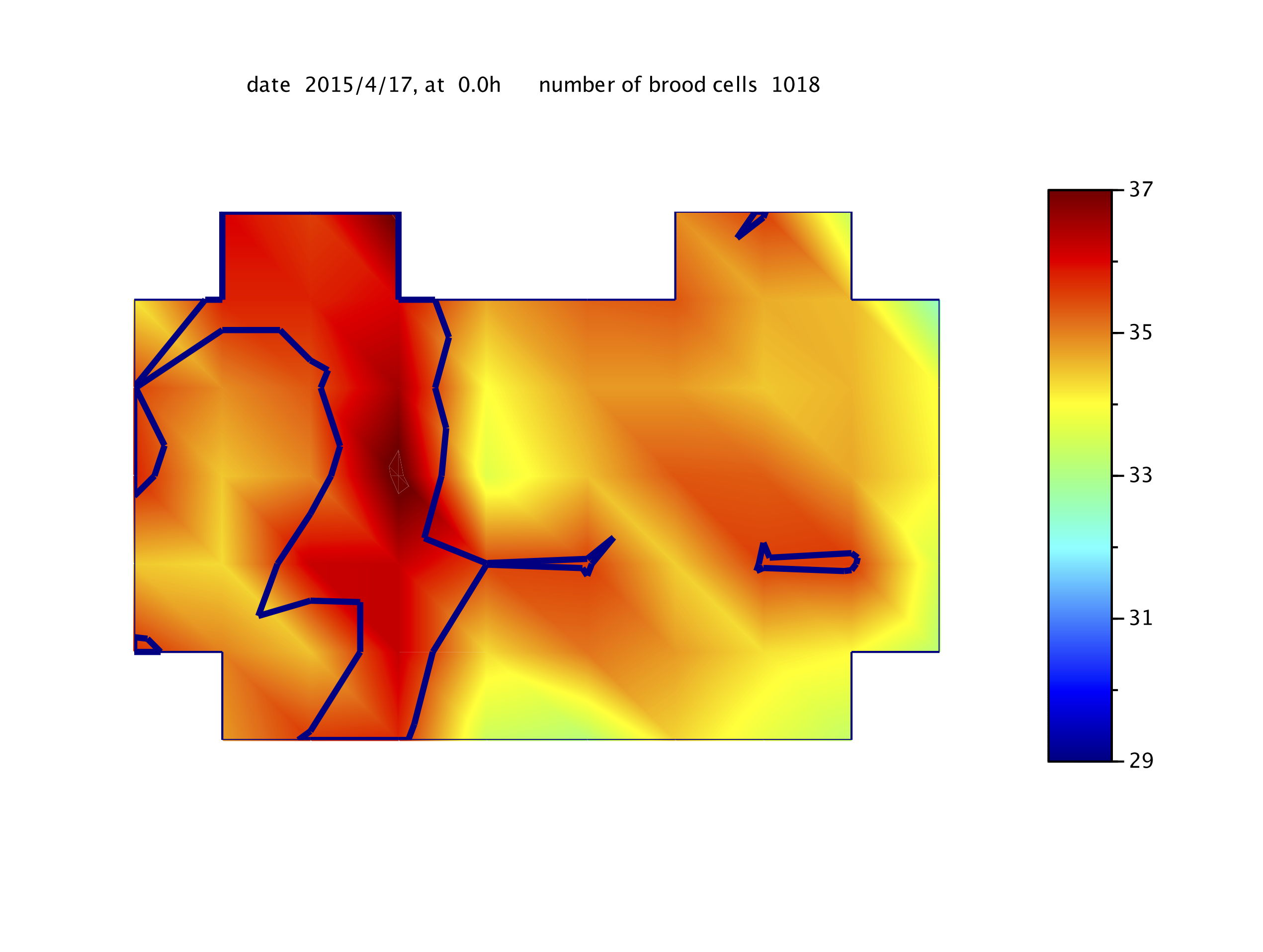}
\caption{{\bf Isotherms for frame 2.}}
\label{iso_50}
\end{center}
\end{figure}
\begin{figure}[H]
\begin{center}
\includegraphics[width=12cm]{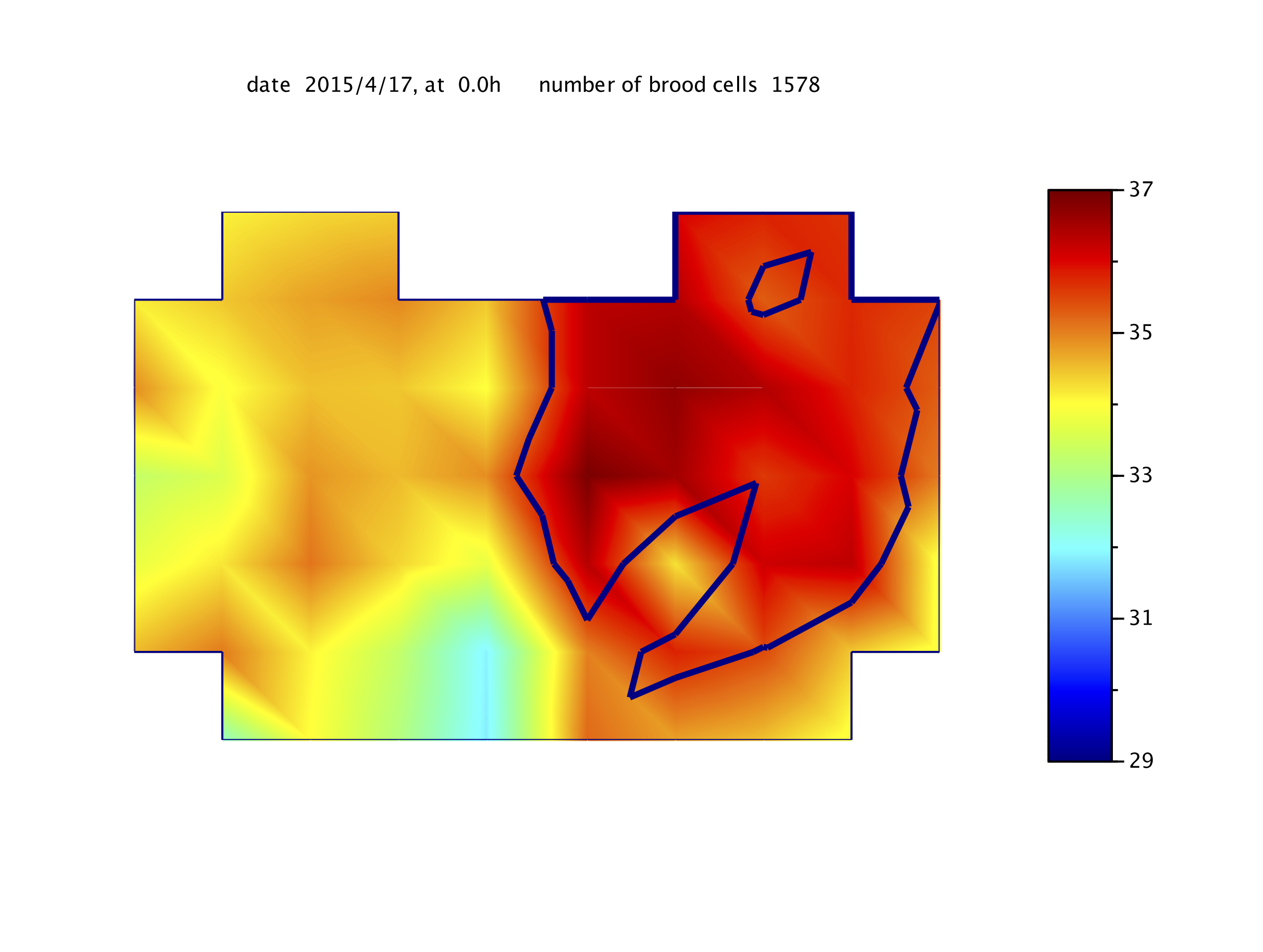}
\caption{{\bf Isotherms for frame 3.}}
\label{iso_30}
\end{center}
\end{figure}
\begin{figure}[H]
\begin{center}
\includegraphics[width=12cm]{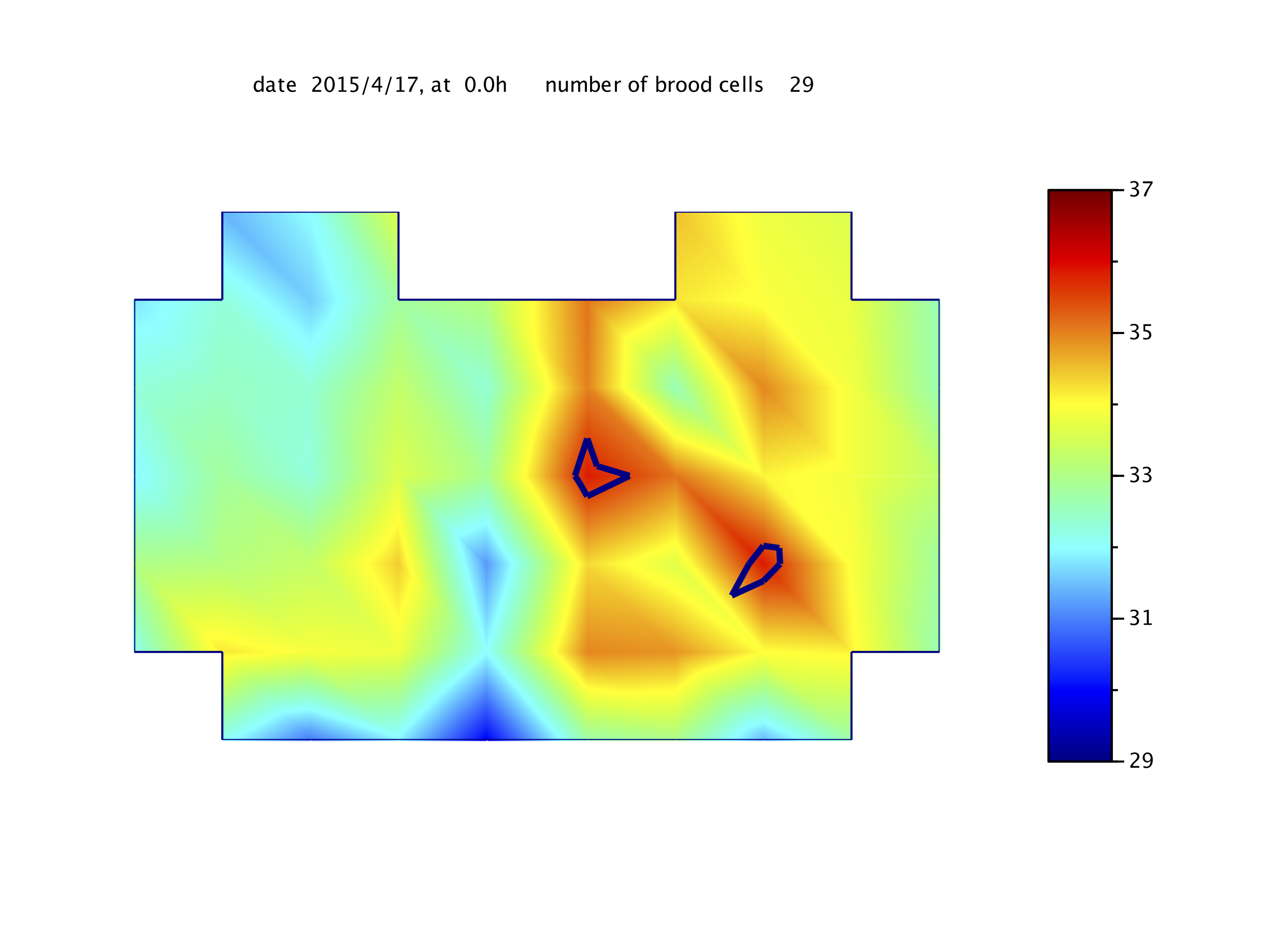}
\caption{{\bf Isotherms for frame 4.}}
\label{iso_80}
\end{center}
\end{figure}
\begin{figure}[H]
\begin{center}
\includegraphics[width=12cm]{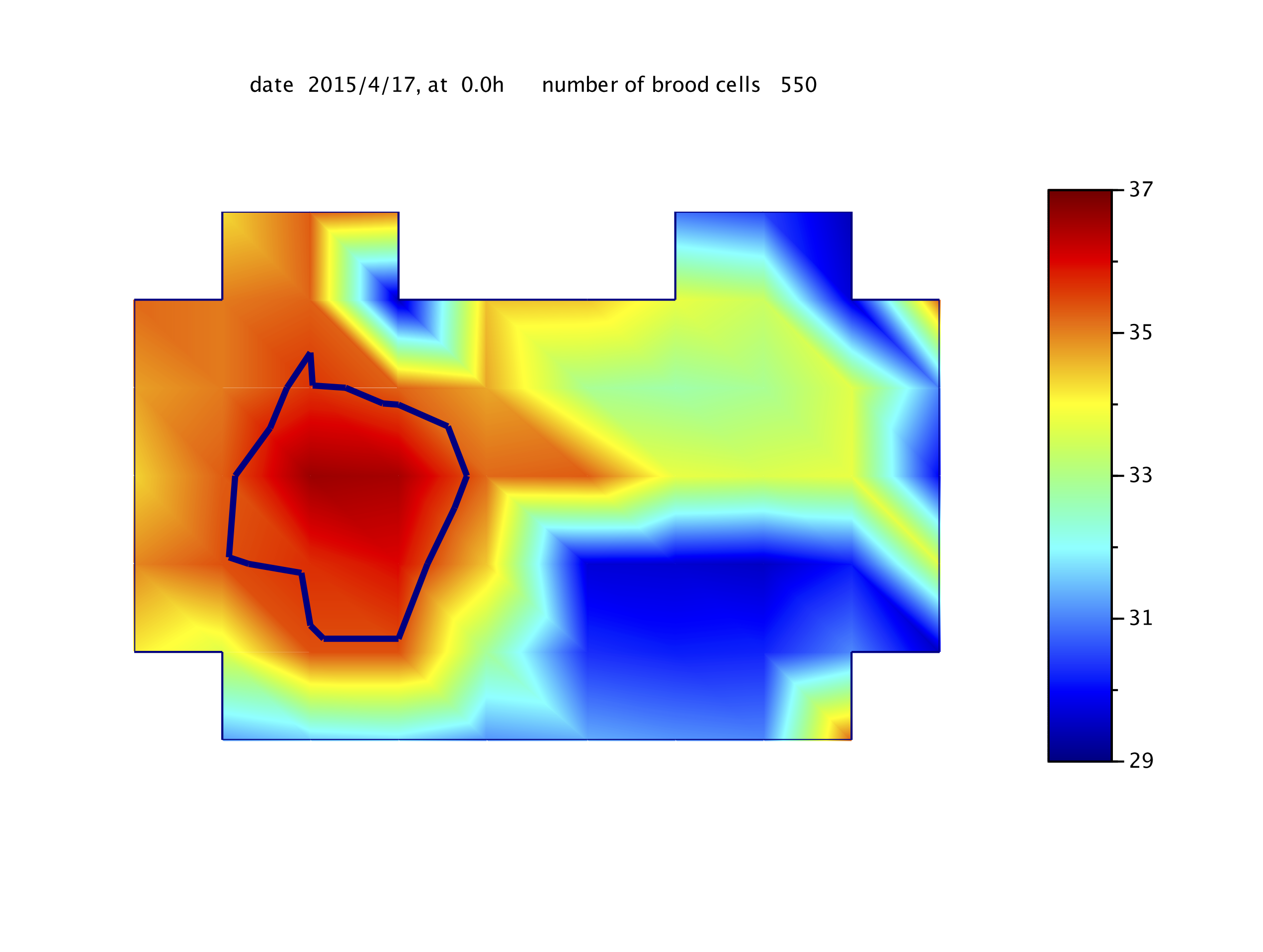}
\caption{{\bf Isotherms for frame 5.}}
\label{iso_40}
\end{center}
\end{figure}

\section{Conclusion}

A new method to continuously monitor the temperature on the frames in a bee hive has been presented. 
The fine mesh on each frame gives a precise 3D representation of the temperature in the hive.

The main advantages of this Bee Cluster 3D system are the following.
\begin{itemize}
\item The system is non-intrusive in the sense that it is not necessary to open the hive to perform measurements.
\item The system is  non-invasive since the sensors merge into the the wax frame.
\item The information is transmitted in real time.
\item The system is energy self-sufficient and is designed for field use. 
\end{itemize}

The presented innovation covers the development of a multi-sensor modular system integrated into 
a ”plug and play” architecture.

It can be used by any beekeeper because
the equipped frames can be manipulated as normal.

Our results show that in a small region of the hive, the temperature remains above 30$^\circ$C throughout the winter,
even if the external temperature is below 0$^\circ$C. The queen is located in this region.

The  Bee Cluster 3D system also provides very useful information concerning the development of the brood nest.
\bigskip

The authors declare no conflicts of interest.
This research did not receive any specific grant from funding agencies in the public, commercial, or not-for-profit sectors.

%

\end{document}